\documentclass{styles/svproc}
\usepackage{url}

\usepackage{graphicx}
\usepackage[subrefformat=parens]{subcaption}
\usepackage{array}

\begin{document}
\mainmatter              
%
\title{Contrastive Cascade Graph Learning\\ for Classifying Real and Synthetic\\ Information Diffusion Patterns}
\titlerunning{CCGL for Classifying Information Diffusion Patterns}  
%
\author{Naoki Shibao\inst{1} \and Sho Tsugawa\inst{2}}
\authorrunning{Naoki Shibao et al.} 
%
\tocauthor{Naoki Shibao, Sho Tsugawa}
\institute{Graduate Schoool of Science and Technology,
  University of Tsukuba\hskip1em\\
  1--1--1 Tennodai, Tsukuba, Ibaraki,
  305--8573 Japan\\
\email{shibao\_n@snlab.tsukuba.ac.jp}
\and
Institute of Systems and Information Engineering,
  University of Tsukuba\hskip1em\\
  1--1--1 Tennodai, Tsukuba, Ibaraki,
  305--8573 Japan\\
\email{s-tugawa@cs.tsukuba.ac.jp}}

\maketitle              

\begin{abstract}
A wide variety of information is disseminated through social media, and content that spreads at scale can have tangible effects on the real world. To curb the spread of harmful content and promote the dissemination of reliable information, research on cascade graph mining has attracted increasing attention.
A promising approach in this area is Contrastive Cascade Graph Learning (CCGL).
One important task in cascade graph mining is cascade classification, which involves categorizing cascade graphs based on their structural characteristics.
Although CCGL is expected to be effective for this task, its performance has not yet been thoroughly evaluated. This study aims to investigate the effectiveness of CCGL for cascade classification.
Our findings demonstrate the strong performance of CCGL in capturing platform- and model-specific structural patterns in cascade graphs, highlighting its potential for a range of downstream information diffusion analysis tasks.
\keywords{Social Media, Cascade Graph, Cascade Classification}
\end{abstract}
\section{Introduction}

A vast amount of information is disseminated daily on social media, and content that spreads widely can significantly influence real-world society. For instance, during the COVID-19 pandemic, numerous rumors, including inaccurate and misleading information, circulated extensively~\cite{covid-bmc}.

In response to the societal impacts of information diffusion on social media, many cascade graph mining methods have been developed to extract valuable insights from the diffusion histories of online content~\cite{cas-1}\cite{cas-2}\cite{cas-3}\cite{cas-4}.
Among the various approaches in this domain, Contrastive Cascade Graph Learning (CCGL)~\cite{ccgl} has recently attracted considerable attention.
CCGL is designed to learn robust graph representations from large-scale unlabeled data, thereby addressing one of the key challenges in cascade graph mining: the limited availability of labeled datasets.

CCGL has demonstrated promising performance in several prediction tasks, most notably in popularity prediction, which forecasts the final size of an information cascade. Other important applications include outbreak prediction, diffusion time estimation, and cascade classification—the task of categorizing information cascades based on their content. While CCGL’s potential has been established for some of these tasks, its effectiveness in cascade classification remains underexplored.

This study focuses on evaluating the effectiveness of CCGL for the task of cascade classification, which is equally important as popularity prediction.
While it has become increasingly evident that the propagation patterns of fake news exhibit distinct structural characteristics~\cite{struct}, research on methodologies that classify cascades based solely on network structural features remains limited.
Developing a classification method grounded in structural information would substantially advance efforts toward the early detection of fake news.
The research questions to be addressed in this paper are as follows:
\setlength{\leftmargini}{30pt}
\begin{itemize}
    \item[\textbf{(RQ1)}] Can the CCGL framework effectively classify information cascades by leveraging both temporal and structural features of cascade graphs?
    \item[\textbf{(RQ2)}] Does CCGL achieve superior classification performance compared to conventional baseline models across varied network structures and diffusion scenarios?
    \item[\textbf{(RQ3)}] How does the quantity of labeled data influence the classification accuracy of the CCGL framework?
\end{itemize}
To answer these questions, we conduct experiments to classify the source datasets of given cascade graphs, using a synthetic dataset created with a network model and an information diffusion model, as well as a real dataset that includes cascades across diverse platforms and languages.
This experimental setup allows us to assess whether CCGL can capture and differentiate structural variations in cascade graphs that arise from differences in model, platform and language.

Our main contributions are summarized as follows.
\begin{itemize}
    \item We demonstrate that cascade graphs can be effectively classified using both temporal and structural features.
    \item We show that CCGL outperforms baseline methods including Random Forest, LightGBM, and Graph Convolutional Network (GCN) in cascade classification tasks.
    \item We find that the accuracy of cascade classification is maintained even when the amount of labeled data is substantially reduced.
\end{itemize}

\section{Experimental Methodology}\label{sec:problem}

\subsection{Problem Formulation}

The cascade classification problem addressed in this study aims to predict the dataset from which a given cascade graph belongs.
Let us consider a piece of information $I$, such as a post on social media or a scholarly article, that is initially published at time $t_{0}$, and subsequently diffused $M$ times through reposts or citations within a time window $t$. This sequence of diffusion events is referred to as a cascade $C$. Here, $t_{j}$ denotes the timestamp at which user $u_{j}$ propagated or reposted the information $I$. The initial posting time by the original user $u_{0}$ is normalized as $t_{0}=0$.
The dataset includes information about diffusion paths, where the diffusion path $e_{k}$ to user $u_{k}$ can be defined as $e_{k}=\lbrace (u_{j}, u_{k}) \vert j,k \in \lbrack 0, M \rbrack, j \neq k \rbrace$.
Let $V$ represent the set of users involved in the cascade $C_{I}(t)$, and let $E$ denote the set of links $e_{k}$, which represent the diffusion paths between users. The cascade graph for information $I$ at time $t$ is then defined as $G_{I}(t)=(V, E)$.
Given the cascade graph $G_{I}(t)$, the cascade classification task aims to predict the label $L_{I}$, which indicates the dataset from which the cascade originates.

To solve this problem, the training data consists of labeled cascade graphs, for which the dataset is known, and unlabeled cascade graphs, for which the dataset is unknown. Using these training data, a model is trained to predict the label $L_{I}$ of the cascade graph $G_{I}(t)$ based on its structure.

\subsection{Synthetic and Real Cascade Datasets}

To investigate the effectiveness of CCGL in cascade classification, we constructed synthetic cascade datasets by simulating information diffusion processes. The procedure consists of two main stages: generating network structures using network models, and simulating information diffusion on these networks using diffusion models.

For network generation, we employed three representative models with distinct structural characteristics: the Barabási–Albert (BA) model~\cite{ba}, the Watts–\\Strogatz (WS) model~\cite{ws}, and the Lancichinetti–Fortunato–Radicchi (LFR) benchmark model~\cite{lfr}.
All models were configured to produce networks with 5,000 nodes, ensuring comparable basic properties while preserving structural diversity.
In the BA model, each newly added node was connected to ten existing nodes. In the WS model, the average degree was set to ten, and the rewiring probability was set to 0.1. For the LFR model, the parameters were as follows: power-law exponent for the degree distribution was set to 2.5, power-law exponent for community size to 1.5, mixing parameter to 0.1, average degree to ten, maximum degree to 100, minimum community size to 100, maximum community size to 600, and the maximum number of iterations to 1,000.

To simulate information diffusion, we used three different models: the Independent Cascade (IC) model~\cite{ic_lt}, the Linear Threshold (LT) model~\cite{ic_lt}, and the Profile model~\cite{profile}. 
For the diffusion parameters, the diffusion probability in the IC model was set to 0.1, the activation threshold in the LT model to 0.09, and the influence strength in the Profile model to 0.3. These values were applied uniformly across all nodes. For each diffusion process, the initial seed node was randomly selected from the entire node set.

By combining the three network models with the three diffusion models, we generated a total of nine distinct synthetic datasets. To enhance the interpretability of cascade structures, we excluded cascades involving fewer than 50 activation events and truncated those exceeding 500 events to retain only the first 500 activations. Each dataset contains exactly 5,000 cascade graphs.

We also use real cascade datasets.
We utilize not only the datasets employed in the original CCGL paper~\cite{ccgl} but also additional datasets used in prior studies that evaluated the effectiveness of CCGL for popularity prediction tasks~\cite{Suzuki}.
The datasets contain cascade data from diverse platforms, languages, and diffusion mechanisms. We investigate whether CCGL can effectively learn the structural differences in cascade graphs that arise from such variations.

\subsection{Training and Evaluation}
To perform cascade classification, we first constructed groups of the datasets by combining multiple data sources. For the synthetic data, we sampled from three datasets generated using different information diffusion models applied to the same underlying network model. Specifically, we selected cascades from each diffusion model in equal proportions, resulting in a dataset containing 6,000 cascades. This process was repeated for each network model, yielding three distinct datasets, referred to as the BA group, WS group, and LFR group.
Similarly, we created three additional datasets based on the same diffusion model applied to different network models. These datasets were labeled as the IC group, LT group, and Profile group.

For the real-world data, we utilized the eleven datasets, categorizing them into three groups based on the average cascade size and platform characteristics: Large, Medium, and Small. 
Specifically, datasets with an average cascade size of 100 or more were assigned to the Large group, those with averages between 50 and 100 to the Medium group, and those with averages below 50 to the Small group.
For each group, we constructed a dataset comprising 4,000 cascade graphs, sampled equally from the original datasets.

For each constructed group, a classification model was trained to predict the source dataset of a given cascade graph. The data was divided such that 60\% of the cascades were used for training and 40\% for testing. The training dataset was split 1:5 for validation and training.
Hyperparameter settings were adopted from the original CCGL study~\cite{ccgl}, unless otherwise noted. Cascade graphs were constructed based on events that occurred within either the first 100 diffusion steps or within one year from the cascade origin, whichever condition was met first.

Finally, the performance of the trained model was evaluated on the test set using the F1 score.
To provide a balanced assessment across all classes, the macro-averaged F1 score was used as the primary evaluation metric.

\subsection{Baselines}

To compare the prediction performance of CCGL, we employed the three baseline models, Random Forest~\cite{randomforest}, LightGBM~\cite{lightgbm}, and GCN~\cite{gcn}.
Each baseline model was trained using the same labeled training data as used for CCGL.
For the Random Forest and LightGBM models, the input features were derived from the cascade graph and included average degree, average path length, link density, and clustering coefficient.
For Random Forest, the loss function was based on Shannon entropy, and the number of decision trees was set to 100.
For LightGBM, the loss function used was Shannon entropy.
The number of boosting iterations was set to 1,000, and early stopping was applied if the loss did not improve for 10 consecutive iterations.
Node features used in the GCN included node degree, average shortest path length, and the timestamp at which the diffusion occurred. We constructed a three-layer GCN with a hidden layer size of 18, a batch size of 5, and 20 training epochs. The Adam optimizer was used with a learning rate of 0.01.

\section{Results}\label{sec:res}
\subsection{Comparison with Baseline}\label{subsec:comp}

We begin by evaluating the performance of CCGL on the cascade classification task using synthetic datasets, comparing it against baseline methods.
The results of the diffusion model classification experiments are summarized in Table~\ref{tab:syn-diff}. For each network model (BA, WS, and LFR), classification models were trained to distinguish among cascades generated using the IC, LT, and Profile models.
As shown in Table~\ref{tab:syn-diff}, CCGL consistently achieved the highest classification accuracy across all network groups.
Although baseline methods performed competitively in the WS and LFR groups, CCGL substantially outperformed them in the BA group—exceeding baseline F1 Score by more than 0.25 in some cases.

\begin{table}[tbp]
    \centering
    \caption{F1 Score for each model of diffusion model classification using synthetic data}
    \begin{tabular}{c|>{\centering}p{7em}|>{\centering}p{5em}|>{\centering}p{3em}|c}
        \hline
         Group &  Random Forest & LightGBM & GCN & CCGL\\
        \hline
        BA Group & 0.70 & 0.71 & 0.61 & 0.96 \\
        WS Group & 0.93 & 0.92 & 0.90 & 0.96 \\
        LFR Group & 0.79 & 0.80 & 0.90 & 0.96 \\
        \hline
    \end{tabular}
    \label{tab:syn-diff}
\end{table}

Table~\ref{tab:syn-net} presents the results for the network model classification task.
In this experiment, we used cascades generated with the same diffusion model (IC, LT, or Profile) across different network topologies (BA, WS, and LFR) and evaluated whether the model could identify the originating network structure.
As seen in Table \ref{tab:syn-net}, overall classification performance for the network model task was lower than that of the diffusion model task.
This is likely due to the fact that diffusion models have a greater influence on the resulting cascade structure than network models, as the diffusion process is applied on top of the network.
Despite this challenge, CCGL still achieved the highest accuracy in all scenarios.

\begin{table}[tbp]
    \centering
    \caption{F1 Score for each model of network model classification using synthetic data}
    \begin{tabular}{c|>{\centering}p{7em}|>{\centering}p{5em}|>{\centering}p{3em}|c}
        \hline
         Group & Random Forest & LightGBM & GCN & CCGL\\
        \hline
        IC Group & 0.70 & 0.71 & 0.68 & 0.71 \\
        LT Group & 0.56 & 0.59 & 0.73 & 0.74 \\
        Profile Group & 0.90 & 0.91 & 0.90 & 0.94 \\
        \hline
    \end{tabular}
    \label{tab:syn-net}
\end{table}

In order to evaluate the effectiveness of CCGL in a real-world setting, a comparative analysis of its predictive performance with baseline models is conducted using real-world datasets.
Table \ref{tab:eva} presents the prediction accuracies of the constructed models.
As shown in the Table \ref{tab:eva}, CCGL consistently outperforms all baseline models in all groups.
In most groups, the differences from the baseline model are larger than in experiments with synthetic data, suggesting that experiments with real-world data may be more challenging than those with synthetic data.
Despite these challenges, CCGL maintained strong and stable classification accuracy.

\begin{table}[tbp]
    \centering
    \caption{F1 Score for each model of classification using real-world data}
    \begin{tabular}{c|>{\centering}p{7em}|>{\centering}p{5em}|>{\centering}p{3em}|c}
        \hline
         Group & Random Forest & LightGBM & GCN & CCGL\\
        \hline
        Small Group & 0.60 & 0.58 & 0.62 & 0.72 \\
        Medium Group & 0.52 & 0.53 & 0.63 & 0.77 \\
        Large Group & 0.78 & 0.77 & 0.60 & 0.90 \\
        \hline
    \end{tabular}
    \label{tab:eva}
\end{table}

As demonstrated by the above experiments, CCGL attains superior classification accuracy in comparison to the baseline models both for synthetic and real datasets.
This suggests that CCGL can effectively learn cascade graph representations that incorporate structural and temporal features.

\subsection{Impact of Training Data Size on CCGL Performance}

Finally, we evaluate the impact of training data size on the prediction performance of CCGL. The original CCGL paper~\cite{ccgl} reported that the model maintains high accuracy even when only limited amounts of labeled data are available. To verify whether this characteristic also holds in the cascade classification task, we conducted experiments using the Medium and Small group datasets from Section \ref{subsec:comp} under reduced training data conditions.
Specifically, we trained models using 10\%, 20\%, 50\%, and 100\% of the training data employed in Section \ref{subsec:comp}, and evaluated their performance accordingly.
The reduced dataset were used in both the fine-tuning and distillation phases.
During the pre-training phase, one of three settings was employed: using only the reduced training data, combining the reduced training data with 4,000 unlabeled cascades from the Large group, or using only the 4,000 cascades from the Large group.

\begin{figure}[tbp]
    \centering
    \begin{tabular}{cc}
        \begin{minipage}[t]{0.45\hsize}
            \centering
            \includegraphics[scale=0.42]{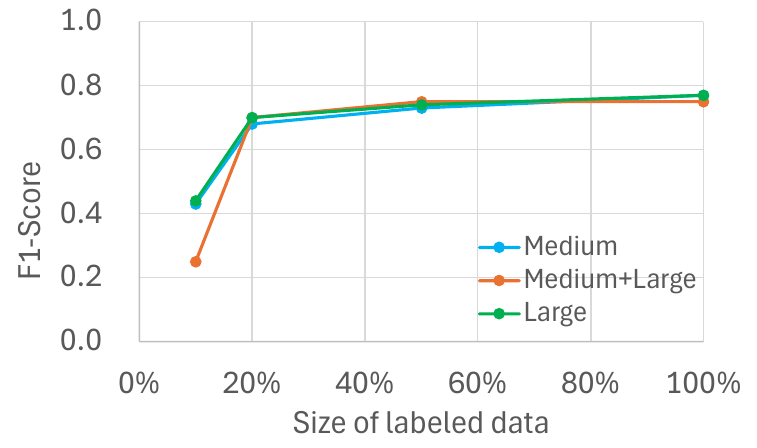}
            \subcaption{Medium Group}
            \label{fig:l2m-F1}
        \end{minipage}&
    
        \begin{minipage}[t]{0.45\hsize}
            \centering
            \includegraphics[scale=0.42]{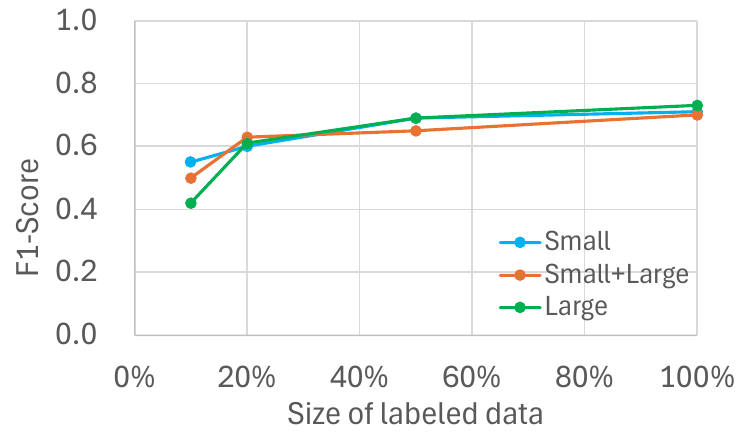}
            \subcaption{Small Group}
            \label{fig:l2s-F1}
        \end{minipage}
    \end{tabular}
    \caption{Change in F1 Score for various datasets used for pre-training with less labeled training data}
    \label{fig:F1}
\end{figure}

Figure \ref{fig:F1} presents the changes in F1 Score observed for the Medium and Small groups under varying training data sizes. The figure compares the outcomes across the three pretraining strategies mentioned above.
As shown in Figure \ref{fig:F1}, the F1 Score remains largely stable when the size of training data is reduced from 100\% to 20\%. In contrast, a significant performance decline is observed when the training size is reduced to 10\%. These results suggest that for the cascade classification task, CCGL maintains robust performance with a moderate amount of labeled data, but its effectiveness deteriorates substantially under extreme data scarcity.

Furthermore, the prediction performance does not exhibit substantial variation across the three pretraining configurations.
Although we hypothesized that leveraging large-scale unlabeled cascade data from other groups in the pretraining phase would help mitigate performance degradation under limited labeled data scenarios, the experimental results indicate that using cascades from the Large group had minimal impact in this context.

\section{Conclusion}\label{sec:conc}

In this study, we evaluated the effectiveness of CCGL for cascade classification, a task that had not been extensively explored in prior work. Through experiments on both synthetic and real-world datasets, we demonstrated that CCGL can accurately classify cascades based solely on structural information, outperforming traditional baselines across diverse platforms. Moreover, the model maintained high performance with as little as 20\% of labeled data, though a sharp drop was observed at 10\%, highlighting a threshold for data efficiency. Our findings extend the applicability of CCGL beyond popularity prediction, positioning it as a robust framework for structure-based cascade analysis in scenarios where content or user metadata is limited.

\section*{Acknowledgments}
 This work was supported by JSPS KAKENHI Grant No. JP25K03105 and the 
Telecommunications Advancement Foundation (TAF).
%
%
%
\bibliographystyle{styles/bibtex/splncs03_unsrt}

\bibliography{myrefs}

\end{document}